\documentstyle[twoside,fleqn,espcrc2]{article}

\input psfig
\pssilent

\def\case#1/#2{{\textstyle\frac{#1}{#2}}}

\def\Tr{\mathop{\rm Tr}\nolimits}

\def\kappa{\varkappa}
\def\kbaux#1{{\kern#1em\Bar{\kern-#1em\varkappa}}}

\def\line#1{\hbox to \hsize {#1}}
\def\centerline#1{\hbox to \hsize {\hss #1\hss}}
\def\Bar#1{\hbox{\boldmath $\bar{\hbox{\unboldmath $#1$}}$}}
\let\tableline\hline
\makeatletter
\def\eqnarray{\stepcounter{equation}\let\@currentlabel=\theequation
\global\@eqnswtrue
\global\@eqcnt\z@\let\\=\@eqncr
\abovedisplayskip\topsep\ifvmode\advance\abovedisplayskip\partopsep\fi
\belowdisplayskip\abovedisplayskip
\belowdisplayshortskip\abovedisplayskip
\abovedisplayshortskip\abovedisplayskip
$$\m@th\halign
to\linewidth\bgroup\@eqnsel\hskip\@centering$\displaystyle\tabskip\z@
  {##}$&\global\@eqcnt\@ne \hfil${##}$\hfil
  &\global\@eqcnt\tw@ $\displaystyle{##}$\hfil
   \tabskip\@centering&\llap{##}\tabskip\z@\cr}
\def\endeqnarray{\@@eqncr\egroup
      \global\advance\c@equation\m@ne$$\global\@ignoretrue
      }
\makeatother

\title{The large-$N$ phase transition of lattice ${\rm SU}(N)$ gauge theories}
\author{Massimo Campostrini\address{Dipartimento di Fisica
dell'Universit\`a and I.N.F.N., I-56126 Pisa, Italy}}
 
\begin{document}

\begin{abstract}
We investigate the large-$N$ phase transition of lattice 
${\rm SU}(N)$ gauge theories in the Wilson formulation,
by performing a Monte Carlo simulation of the twisted Eguchi-Kawai
model.  A variant of the multicanonical algorithm allows a detailed
exploration of the phase transition and a precise determination of
the transition temperature.
\end{abstract}
\maketitle

\section{Introduction}

The large-$N$ limit of ${\rm SU}(N)$ gauge theories is of considerable
interest from the phenomenological point of view, and is one of our
sources of understanding of non-perturbative QCD.  The theoretical
side is no less interesting, allowing e.g.\ to establish precise
relationships between Yang-Mills theories and string theories.

A crucial property of large-$N$ field theories is 
{\em factorization\/}: connected Green's functions of invariant
quantities are suppressed with respect to the corresponding
disconnected parts by powers of $1/N$.  One of the most notable
consequences of factorization is the possibility of constructing 
{\em reduced\/} models, i.e.\ single-site models which reproduce a
lattice gauge theory in the $N\to\infty$ limit
\cite{Das-review,Rossi-Campostrini-Vicari}.

Despite the considerable simplifications occurring for large $N$, many
interesting models have not been solved analytically; therefore we
must resort to approximate techniques, such as numerical simulations.

\section{The TEK model}

The most promising reduced version of the ${\rm SU}(N)$ lattice gauge
theory (in the Wilson formulation) is the {\em twisted Eguchi-Kawai\/}
model (TEK) \cite{GonzalezArroyo-Okawa-twisted}.  Let us define a set
of 4 traceless ${\rm SU}(N)$ matrices $\Gamma_\mu$ obeying 't Hooft
algebra
\[
\Gamma_\nu \Gamma_\mu = Z_{\mu\nu}\Gamma_\mu \Gamma_\nu, \qquad
Z_{\mu\nu} = \exp\!\left({2i\pi\over N} n_{\mu\nu}\right),
\]
where $n_{\mu\nu}$ is an antisymmetric integer-valued tensor.
$\Gamma_\mu$ are the matrices implementing the translations by one
lattice spacing in the $\mu$ direction. The observables of the reduced
model are obtained by applying the reduction prescription
\[
U_\mu(x) \to T(x)U_\mu T(x)^\dagger, \qquad
T(x)= \prod_\mu (\Gamma_\mu)^{x_\mu}
\]
to the corresponding quantity of the full model.  The action of the
TEK model is obtained by reduction of the Wilson action:
\begin{eqnarray*}
&&\beta N^2\,E(U) \equiv S_{\rm TEK}(U) = \\
&&\qquad N\beta \sum_{\mu>\nu}
\Tr\left[ Z_{\mu\nu}U_\mu U_\nu U_\mu^\dagger U_\nu^\dagger 
+ {\rm h.c.}\right].
\end{eqnarray*}

For a proper choice of $\Gamma_\mu$, the Schwinger-Dyson equations of
the TEK model reproduce in the large-$N$ limit the Schwinger-Dyson
equations of the Wilson lattice gauge theory.  We adopt the choice of
Ref.~\cite{GonzalezArroyo-Okawa-EK}: $N$ is constrained to be a
perfect square, $N=L^2$, and $n_{\nu\mu}=L$ for all $\nu>\mu$.  $L$
takes the r\^ole of the lattice size, and the number of degrees of
freedom is proportional to $L^4$.

The TEK model develops a first-order phase transition for
$N\to\infty$.  It corresponds to the large-$N$ limit of the
first-order phase transition of ${\rm SU}(N)$ lattice gauge theories.
We will study the phase transition of the TEK model by Monte Carlo
simulation.

\section{The multicanonical algorithm}

The choice of updating algorithm is extremely important in the
neighborhood of a phase transition.  Most numerical work on the TEK
model adopted the heat-bath updating algorithm (HB) devised in
Ref.~\cite{Fabricius-Haan}.  The HB algorithm is generally quite
efficient, but near the phase transition it is plagued by the familiar
{\it supercritical slowing down\/}; according to the experience
accumulated in the last decade, this problem can be solved by devising 
a suitable {\it multicanonical\/} algorithm \cite{Berg-Neuhaus}.

Since we are studying a temperature-driven phase transition, it is
natural to choose a reweighting function depending only on the energy:
we generate configurations according to the probability distribution
\[
P(\beta,U) \propto w(E(U))\,\exp(-\beta N^2 E(U))
\]
and compute the canonical expectation value of an observable as
\[
\left<{\cal O}\right> = \sum_c w^{-1}(E(c))\,{\cal O}(c) \Bigm/
\sum_c w^{-1}(E(c)).
\]

If we can find a is reasonably accurate ansatz $\bar\rho$ to the energy
distribution $\rho(E)$,
we can construct the reweighting function
\begin{eqnarray}
&&w(E) = 1/\bar\rho(E_-), \quad E \le E_-, \nonumber \\
&&w(E) = 1/\bar\rho(E),\hphantom{{}_+} \quad
         E_- \le E \le E_+, \nonumber \\
&&w(E) = 1/\bar\rho(E_+), \quad E_+ \le E,
\label{w}
\end{eqnarray}
which flattens the probability distribution between the two peaks, and
we can obtain a reasonable tunneling rate between the two vacua.  The
resulting multicanonical distribution can be simulated using an
efficient Metropolis procedure.

Our first ansatz was
\begin{eqnarray}
&&\rho_{\rm BL}(E) \equiv a_+ g_+(E) + a_- g_-(E) +
\gamma\,\theta(E-E_+) \nonumber \\
&&\qquad\times\,\theta(E_--E)\,
\left(1 - g_+(E)\right)\,\left(1 - g_-(E)\right), \nonumber \\
&&g_\pm(E) = \exp\!\left[-{(E-E_\pm)^2 \over 2 \sigma_\pm^2}\right],
\label{BL}
\end{eqnarray}
where all the parameters $E_\pm$, $\sigma_\pm$, $a_\pm$, and $\gamma$
are $N$-dependent; the factors of $\left(1 - g_\pm(E)\right)$ ensure
the smoothness of the distribution for $E=E_\pm$.  This is essentially
a Binder-Landau ansatz \cite{Binder-Landau} corrected for mixed-phase
contributions \cite{Billoire}; the resulting algorithm (M1) works
quite well up to $N=25$, but it becomes very inefficient as $N$
increases further.

\begin{figure}[tb]
\vbox to 5cm {
\centerline{\psfig{figure=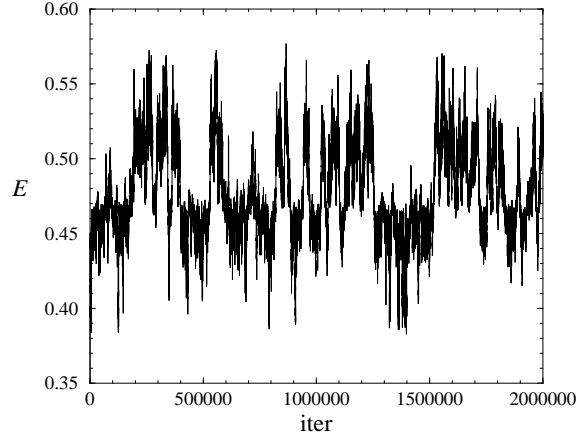,width=7.5cm}}
\vss }
\caption{The Monte Carlo time evolution of the energy, using the M2
algorithm for $N=64$.}
\label{fig:history64}
\end{figure}

It turns out that $\rho$ does not follow the Gaussian behavior of Eq.\ 
(\ref{BL}) in the region $E_- \ll E \ll E_+$; if we start from one
peak and move towards the other, the distribution will eventually
follow an exponential law.  In order to reproduce this behavior, we
introduce two new parameters 
$\bar E_\pm$, $E_- < \bar E_- < \bar E_+ < E_+$,
and replace $g_\pm$ in Eq.~(\ref{BL}) with
\begin{eqnarray}
&&g_\pm(E) = \exp\!\left[-{(E-E_\pm)^2 \over 2 \sigma_\pm^2}\right], \ \ 
E \le \bar E_-\ (E \ge \bar E_+), \nonumber \\
&&g_\pm(E) = \exp(r_\pm E + s_\pm), \ \ 
E \ge \bar E_-\ (E \le \bar E_+),
\label{BLimp}
\end{eqnarray}
where $r_\pm$ and $s_\pm$ are determined by the condition that the two
branches of $g_\pm$ join smoothly (up to the first derivative) for
$E=\bar E_\pm$.  The new algorithm (M2) works remarkably well up to
$N=64$, as shown in Fig.~\ref{fig:history64}.

The main drawback of algorithms M1 and M2 is that the multicanonical
parameters must be tuned to an ever finer degree with increasing $N$.
Thanks to the absence of tunneling in a canonical simulation, $E_\pm$
and $\sigma_\pm$ can be estimated accurately by performing a canonical
simulation starting from a disordered and an ordered configuration
respectively, using the HB algorithm.  The other parameters can be
estimated roughly from the simulations at lower $N$; this estimate
needs to be refined by performing successive multicanonical
simulations and looking at the resulting energy distribution.  For
$N=64$ this required more then 10 simulations at moderate statistics
(about 200k sweeps), with a computational effort comparable to the
high-statistics simulation with the optimized parameters.

\begin{table}[tb]
\setlength{\tabcolsep}{0.85pc}
\caption{Summary of high-statistics simulations.}
\label{tab:summary}
\begin{tabular}{ccccl}
\tableline
$N$ & $\beta$ & alg & stat & 
\multicolumn{1}{c}{$\gamma$}\vphantom{$1^{1^{1^1}}$} \\
\tableline
25 & 0.3580 & HB & 5M & \vphantom{$1^{1^{1^1}}$} \\
25 & 0.3574 & M1 & 5M & $5 \times 10^{-3}$ \\
36 & 0.3585 & M2 & 5M & $1 \times 10^{-5}$ \\
49 & 0.3588 & M2 & 4M & $1 \times 10^{-10}$ \\
64 & 0.3595 & M2 & 2M & $5 \times 10^{-17}$ \\
\tableline
\end{tabular}
\end{table}

\section{Results}

A summary of our high-statistics simulations is presented in
Table~\ref{tab:summary}.

\begin{figure}[b]
\vbox to 5cm {
\centerline{\psfig{figure=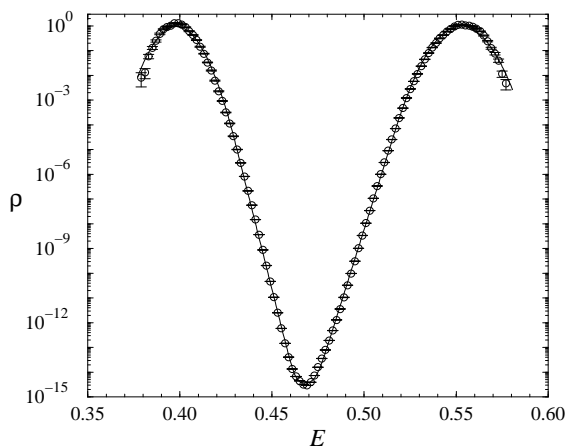,width=7.5cm}}
\vss }
\caption{The (unnormalized) energy distribution $\rho$, compared with
a best fit to Eq.\ (\protect\ref{BLimp}).}
\label{fig:fit64}
\end{figure}

The quality of our ansatz (\ref{BLimp}) can be judged from
Fig.~\ref{fig:fit64}, where the worst case $N=64$ is presented.  The
ansatz is not consistent with the data within the statistical errors,
but it is more then accurate enough for the purpose of optimizing the
multicanonical algorithm.

It is interesting to notice the exponential fall-off of the energy
distribution between the two peaks, which for $N=64$ is followed to
great accuracy for several orders of magnitude.  On the other side of
the two peaks, the energy distribution falls much faster, even faster
then the Gaussian behavior assumed in Eqs.~(\ref{BL}) and
(\ref{BLimp}).

\begin{figure}[tb]
\vbox to 5cm {
\centerline{\psfig{figure=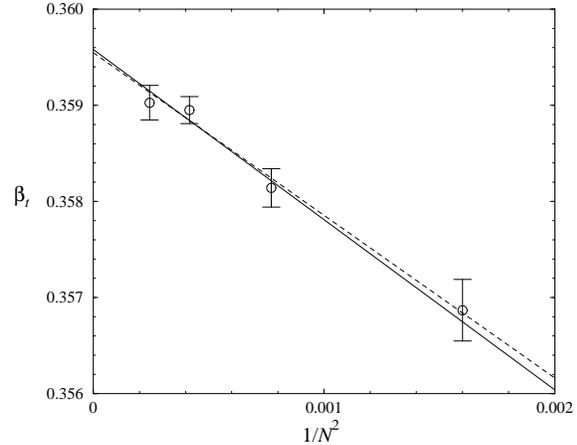,angle=270,width=7.5cm}}
\vss }
\caption{Linear fit in $1/N^2$ to $\beta_t$.  The solid line is a fit
excluding the rightmost point ($N=25$); the dashed line is a fit
including all the four points plotted.}
\label{fig:betac}
\end{figure}

A single multicanonical simulation can be used to obtain results for a
(small) range of $\beta$, using the {\em reweighting\/} technique.
Our estimator of the inverse transition temperature $\beta_t$ is the
value of $\beta$ which maximizies the specific heat.
Fig.~\ref{fig:betac} shows that $\beta_t$ is in perfect agreement with
the linear behavior in $1/N^2$ expected for a first order phase
transition.  Extrapolating to $N=\infty$ by a linear fit, we obtain
\[
\beta_t = 0.3596(2).
\]

\end{document}